\documentclass{iau}

\usepackage{amsmath}
\usepackage{graphicx}
\usepackage{multirow}

\begin{document}

\lefttitle{S. \"{O}zdemir et al.}
\righttitle{Probing the Gaia atmospheric parameters of stars in globular clusters}

\jnlPage{1}{7}
\jnlDoiYr{2025}
\doival{10.1017/xxxxx}
\volno{395}
\pubYr{2025}
\journaltitle{Stellar populations in the Milky Way and beyond}

\aopheadtitle{Proceedings of the IAU Symposium}
\editors{J. Mel\'endez,  C. Chiappini, R. Schiavon \& M. Trevisan, eds.}

\title{Probing the Gaia atmospheric parameters of stars in globular clusters}

\author{S. \"{O}zdemir, J.~E. Mart\'{\i}nez-Fern\'andez \& R. Smiljanic}
\affiliation{Nicolaus Copernicus Astronomical Center, Polish Academy of Sciences, ul. Bartycka 18, 00-716, Warsaw, Poland}

\begin{abstract}
Numerous stellar surveys have been or will provide photometric, astrometric, and spectroscopic data for a large number of stars in the Milky Way and neighbouring galaxies. Modern data processing tools and analysis methods are needed to deal with these data sets and obtain accurate and precise results. In this context, we are developing a new spectroscopic analysis pipeline based on the differential analysis method. The CHEmical Survey analysis System (CHESS) aims to automate the steps needed to obtain high-quality stellar parameters and abundances from large samples of spectra. To automatically identify the spectra of similar stars that are suitable for a differential analysis, CHESS first performs what we call a similarity analysis by directly using the observed spectra. This step of analysis uses unsupervised machine learning algorithms (such as dimensionality reduction methods). To validate the findings, we used atmospheric parameters from several catalogues (in particular those available in Gaia DR3). Alternatively, such a similarity analysis also serves as a consistency check of the atmospheric parameters in these catalogues.  Here, we present our method for finding similar stars in globular clusters and the first preliminary results of their atmospheric parameters. 
\end{abstract}


\maketitle

\section{Introduction}

We are developing a new spectroscopic analysis pipeline, called CHEmical Survey analysis System (CHESS), based on the differential analysis method and assisted by machine learning techniques. As a first application, we will use CHESS to re-analyse the archival spectra obtained with the Ultraviolet and Visual Echelle Spectrograph \citep[UVES,][]{2000SPIE.4008..534D} on the Unit Telescope 2 of the Very Large Telescope (VLT), Paranal Observatory, Chile. The project discussed here focusses on the analysis of globular clusters. However, CHESS will eventually analyse all the spectra of F-, G-, and K-type stars in the archive. See Smiljanic et al. (this volume) for a brief summary of CHESS and its capabilities.

The stars that CHESS identifies as very similar have almost the same values of effective temperature ($\pm$250 K), surface gravity ($\pm$0.2 dex), and metallicity ($\pm$0.2 dex). After that, differential analysis can be applied to analyse the sample compared to similar reference stars that have accurate atmospheric parameters (Martínez-Fernández, Özdemir et al., in prep). Here, we present preliminary results of our approach applied to stars in globular clusters. 

\section{Sample}

The CHESS pipeline is being mainly designed to (re-)analyse processed spectra from the ESO Archive\footnote{\texttt{https://archive.eso.org/cms.html}} that have been taken with UVES, a high-resolution (40\,000 $<$ R $<$ 110\,000) echelle spectrograph that can work from the ultraviolet atmospheric cut-off (about 300 nm) up to 1100 nm. 

To select the spectra for analysis, we used the centre coordinates of all Galactic globular clusters from the 2010 version of the \cite{1996AJ....112.1487H} catalogue. We compiled all UVES archival reduced spectra available within a radius of 20 arcmin around the centre of each cluster. To test the similarity analysis, we removed spectra that had a signal-to-noise ratio (SNR) $<$ 10. We collected a total of 2800 spectra of 1300 stars.

We then applied the following steps to the data: identifying targets, masking telluric affected regions, degrading the spectral resolution to the lowest common value (R $\sim$ 30\,000), calculating and correcting the radial velocity, resampling the spectra to a common solution, normalisation of the continuum. After these steps, we have sets of spectra on the same scale that can be compared in a similarity analysis. In addition, we have atmospheric parameters from the literature that can be used to quantify the success of the analysis.

We also compiled spectra of globular cluster stars obtained with the Gaia Radial Velocity Spectrometer \citep[RVS,][]{2018A&A...616A...5C}. The RVS is a medium-resolution spectrograph (R $\simeq$ 11 500) with high spectral fidelity operating in the wavelength range of 845–872 nm. We used the Harris catalogue as above to select stars, but we also filtered the sample using membership probabilities ($P_{\rm memb} \geq 0.7$) from \cite{2021MNRAS.505.5978V}. Using a SNR $\geq$ 40 criterion, we ended up with combined mean Gaia RVS spectra for 215 globular cluster stars.

\begin{figure}
    \centering
    \includegraphics[scale=0.28]{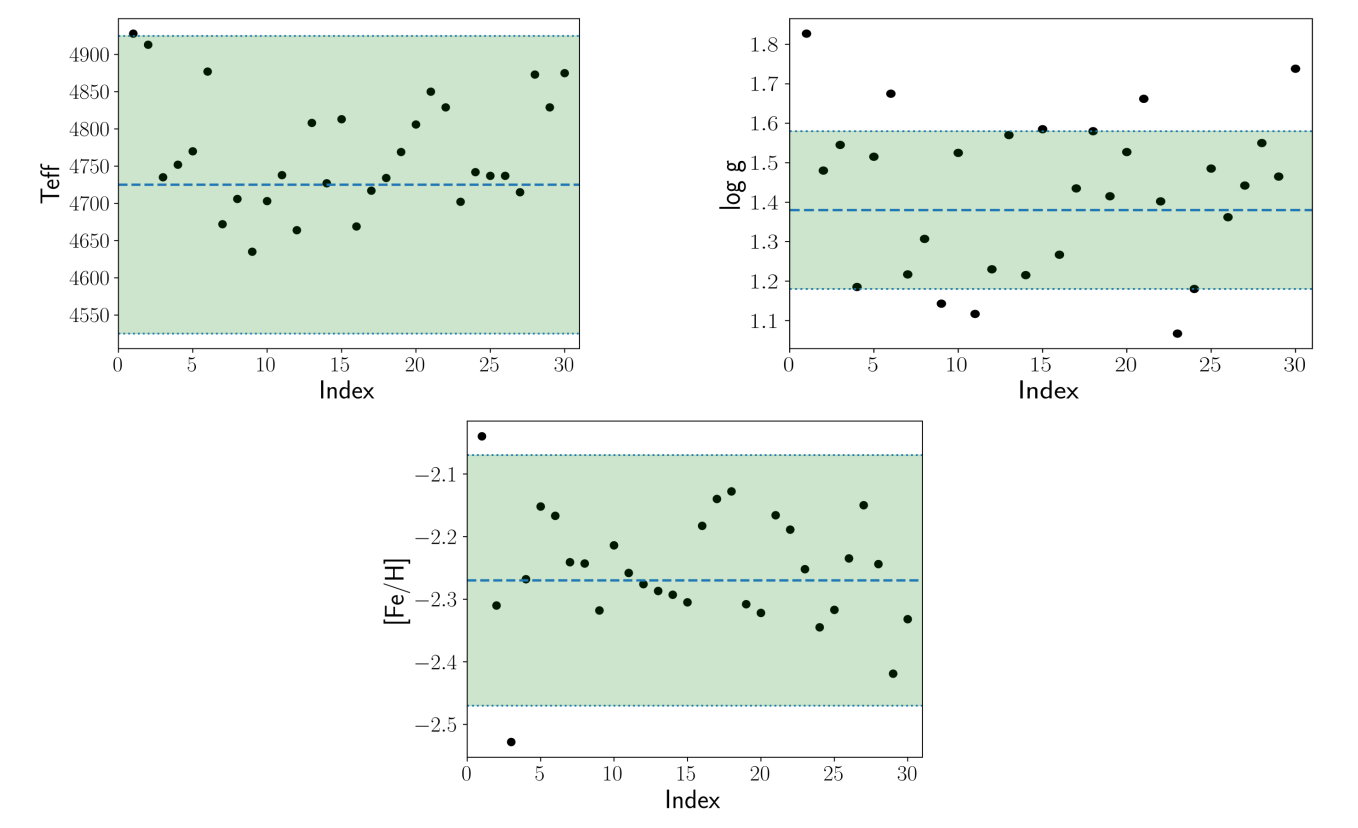}
    \caption{Atmospheric parameters of group of 29 stars found to be similar to BD+09 2870. The dashed lines show the values of BD+09 2870 for the given parameter while the green shaded areas indicate regions of $\pm$200 K for $T_{\rm eff}$ and $\pm$0.2 dex for log g and [Fe/H].}
    \label{fig:teff}
\end{figure}

\section{Similarity analysis}
\subsection{Using the UVES sample}

To find similar stars using directly the homogenised spectra, CHESS uses a dimensionality reduction technique called t-SNE. However, the bi-dimensional projection maps of t-SNE do not have axes that directly correspond to physical quantities. To help in defining a set of similar stars, we defined a metric that is described in Mart\'{\i}nez-Fern\'andez et al. (in preparation). With this metric, we are able to separate groups of stars with atmospheric parameters similar to one of our reference stars without performing a full spectroscopic analysis. In Figure \ref{fig:teff}, we show the results of a differential analysis of atmospheric parameters of 29 stars similar to BD+09 2870 (with reference parameters $T_{\rm eff}$ = 4725 K, log g = 1.38 and [Fe/H] =$-$2.27). Our similarity analysis works very well to identify similar stars only by using their spectra.

\subsection{Using the Gaia RVS sample}

\begin{figure}
    \centering
    \includegraphics[scale=0.55]{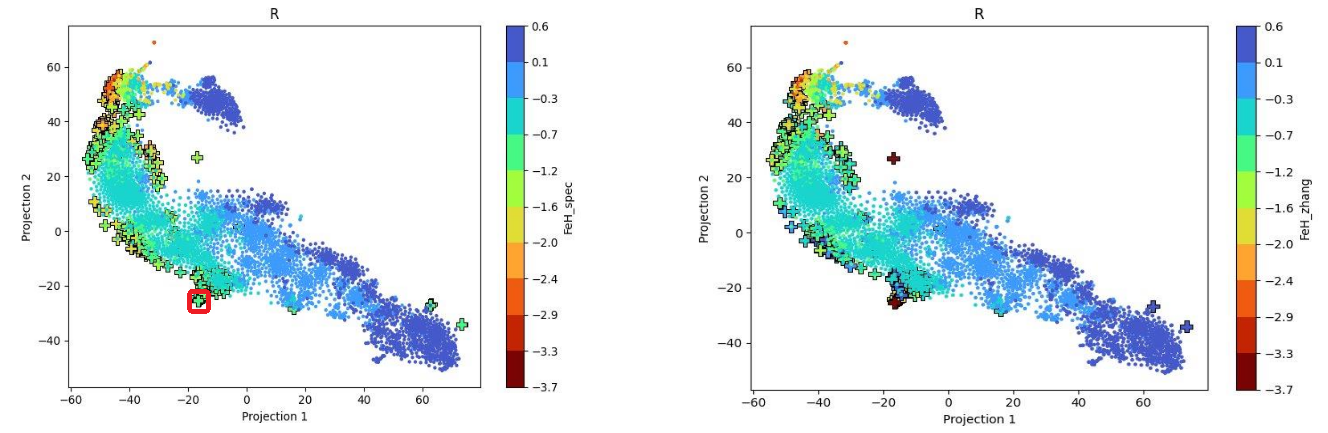}
    \caption{Two-dimensional t-SNE projection maps for the Gaia RVS spectra. Reference stars are shown with circles and globular cluster stars as plus markers. Symbols are colour-coded by the value of [Fe/H] from the Gaia RVS analysis of \citet{2023A&A...674A..29R}, left panel, and from the catalogue of \citet{2023MNRAS.524.1855Z}, right panel. The red square on the left panel shows the location of stars discussed in the text.}
    \label{fig:tsne}
\end{figure}

We also performed the similarity analysis using the t-SNE algorithm on our sample of Gaia RVS spectra (Figure \ref{fig:tsne}). As reference stars in this case, we used a selection of 4200 stars with good quality atmospheric parameters from Gaia DR3. This analysis can highlight differences between catalogues of stellar parameters. An example is a group of spectra (13 stars, inside the red square on the left panel of Fig. \ref{fig:tsne}) which are similar in both visual inspection and Gaia spectroscopic parameters. Further investigation showed that 12 out of 13 stars in that group are members of the globular cluster 47 Tuc. According to \cite{1996AJ....112.1487H}, 47 Tuc is monometallic cluster with [Fe/H] = $-$0.72. This group of stars has an average of [Fe/H] = $-$0.88 ($\pm$0.14) in the Gaia RVS values \citep{2023A&A...674A..29R}. However, the average value of the \cite{2023MNRAS.524.1855Z} catalogue is [Fe/H] = $-$2.58 ($\pm$0.71).

\section{Summary and future steps}
The CHESS pipeline can identify similar stars ($\pm$ 250 K in $T_{\rm eff}$, $\pm$ 0.2 dex in log g and [Fe/H]) just by comparing their spectra (Mart\'{\i}nez-Fern\'andez et al. in prep.). The next step is to perform the  differential analysis to obtain precise and accurate chemical abundances in globular cluster stars. All elements covered by the UVES wavelength coverage will be studied. Our preliminary studies with Gaia RVS data and different catalogues confirm the need for caution when using parameters from large databases (especially metallicity). \\

\noindent \textbf{Acknowledgements}

The authors acknowledge support from the National Science Centre, Poland, project 2019/34/E/ST9/00133.


\begin{thebibliography}{}
\bibitem[Cropper et al.(2018)]{2018A&A...616A...5C} Cropper, M., Katz, D., Sartoretti, P., et al.\ 2018, A\&A, 616, A5;
\bibitem[Dekker et al.(2000)]{2000SPIE.4008..534D} Dekker, H., D'Odorico, S., Kaufer, A., et al.\ 2000, Proceedings of SPIE, 4008, 534;
\bibitem[Harris(1996)]{1996AJ....112.1487H} Harris, W.~E.\ 1996, AJ, 112, 1487;
\bibitem[Recio-Blanco et al.(2023)]{2023A&A...674A..29R} Recio-Blanco, A., de Laverny, P., Palicio, P.~A., et al.\ 2023,{A\&A}, 674, A29;
\bibitem[Vasiliev \& Baumgardt(2021)]{2021MNRAS.505.5978V} Vasiliev, E. \& Baumgardt, H.\ 2021, MNRAS, 505, 5978;
\bibitem[Zhang et al.(2023)]{2023MNRAS.524.1855Z} Zhang, X., Green, G.~M., \& Rix, H.-W.\ 2023, MNRAS, 524, 1855.

\end{thebibliography}
\end{document}